\documentclass[conference]{IEEEtran}
\usepackage{xcolor}
\usepackage{fancyhdr}
\usepackage{tgschola}
\usepackage{lastpage}
\usepackage{amsthm,amssymb,amsmath,mathtools,fixmath}
\usepackage{amsbsy}
\usepackage{moreverb}
\usepackage{amsfonts}
\usepackage{graphicx}
\usepackage{tikz}
\usepackage{listings}
\usepackage{color}
\usepackage{latexsym}
\usepackage{animate}
\colorlet{linkequation}{blue}
\usepackage{lipsum}% http://ctan.org/pkg/lipsum

\usepackage[colorlinks,bookmarksopen,bookmarksnumbered,citecolor=black,urlcolor=black]{hyperref}
\hypersetup{linkcolor=black}
\DeclareMathOperator*{\maxi}{maximize}

\ifCLASSINFOpdf
\else
\fi
\begin{document}
\title{Time-Fractional User Association in \\ Millimeter Wave MIMO Networks}

\author{\IEEEauthorblockN{Alireza Alizadeh and Mai Vu}
\IEEEauthorblockA{Department of Electrical and Computer Engineering,
Tufts University, Medford, USA\\
Email: \{alireza.alizadeh, mai.vu\}@tufts.edu}
%\and
%\IEEEauthorblockN{Mai Vu}
%\IEEEauthorblockA{Department of Electrical and Computer Engineering\\
%Tufts University\\
%Medford, USA\\
%Email: mai.vu@tufts.edu}
}


%Telephone: (800) 555--1212, Fax: (888) 555--1212}


\maketitle
\begin{abstract}
User association determines which base stations a user connects to, hence affecting the amount of network interference and consequently the network throughput. Conventional user association schemes, however, assume that user instantaneous rates are independent of user association. 
In this paper, we introduce a new load-aware user association scheme for millimeter wave (mmWave) MIMO networks which takes into account the dependency of network interference on user association. This consideration is well suited for mmWave communications, where the links are highly directional and vulnerable to small channel variations. 
We formulate our user association problem as a mixed integer nonlinear programming (MINLP) and solve it using the genetic algorithm. We show that the proposed method can improve network performance by moving the traffic of congested base stations to lightly-loaded base stations and adjusting the interference accordingly. Our simulations confirm that our scheme results in a higher network throughput compared to conventional user association techniques.
\end{abstract}


% \fi
%\IEEEpeerreviewmaketitle

\section{Introduction}
Bandwidth shortage is a major challenge for current wireless networks. Most of the available spectrum at microwave frequencies is occupied while there is a pressing need for higher throughputs and larger bandwidths. 
During the past few years, mmWave frequencies have attracted the interest of academia and industry due to the capability of multi-Gbps data rates and the huge amount of bandwidth available at frequencies between 30 - 300 GHz. The mmWave band is a promising candidate for the next generation of cellular networks (5G).

User association plays an important role in the resource allocation for cellular networks. The conventional max-SINR user association is sub-optimal in dense networks as incoming user equipments (UEs) may receive the strongest signal from a congested base station (BS) and overload it. 
In this case, we need to design a load-aware user association scheme in order to move the traffic of congested BS to lightly-loaded smaller BSs.

Load balancing user association schemes are studied for single antenna heterogeneous networks (HetNets) in \cite{Andrew} and for massive MIMO networks in \cite{Caire}, where it is assumed that the user instantaneous rates converge to deterministic values independent of user association and active user sets. This assumption leads to a simple but inaccurate full interference structure that degrades the network sum rate, since the effect of association coefficients on network interference structure is ignored.
%Hence, the network interference is also independent from user association.
Further, the interference which was ignored in 60-GHz mmWave indoor networks \cite{Atha}, is no longer negligible in a dense mmWave outdoor network at frequencies considered for cellular application (28, 38, and 73 GHz) \cite{Niu}.  
A few existing works also considered the joint problem of user association and beamforming design (see \cite{Hong}, \cite{Sanjabi}, and the references therein). This problem is shown to be NP-hard, and researchers usually proposed different iterative algorithms to achieve near-optimal solutions.

In this paper, we formulate and solve an optimization problem for optimal user association in mmWave MIMO networks. The first step is the generation of a mmWave channel which is drastically different from an i.i.d channel. 
The considered channel model is based on the clustered channel model introduced in \cite{SS} and the 3GPP-style 3D mmWave channel proposed in \cite{Nokia}. The next step is to formulate an optimization problem and solve it in order to find the optimal user association. 
We introduce the \textit{Activation Matrix} which defines UE-BS connections in each time slot, from which association coefficients are derived as its time average.
Unlike existing works, here we assume that the user instantaneous rate is a function of user association, as is the case in mmWave. Consequently, the total interference coming from other BSs (while serving other UEs) also depends on association coefficients and has a considerable effect on network sum rate.

\section{Channel and System Model}
\subsection{mmWave Channel Model}
The mmWave channel has completely different characteristics compared to i.i.d. channel. 
The channel model considered in this paper is based on the clustered channel model introduced in \cite{SS} and the 3GPP-style 3D channel model proposed for the urban micro (UMi) environments in \cite{Nokia}, which is developed using a ray-tracing study.
This channel model has $C$ clusters with $L$ rays per cluster, and it can be expressed as
\begin{align}
H=\frac{1}{\sqrt{CL}}\sum_{c=1}^{C}\sum_{l=1}^{L} \sqrt{\gamma_c}~\mathbf{a}(\phi_{c,l}^{\textrm{UE}},\theta_{c,l}^{\textrm{UE}}) ~\mathbf{a}^*(\phi_{c,l}^{\textrm{BS}},\theta_{c,l}^{\textrm{BS}})
\end{align}
where $\gamma_c$ is the gain of the $c$th cluster. The parameters $\phi^{\textrm{UE}}$, $\theta^\textrm{UE}$, $\phi^\textrm{BS}$, $\theta^\textrm{BS}$ represent azimuth angle of arrival (AoA), elevation angle of arrival (EoA), azimuth angle of departure (AoD), and elevation angle of departure (EoD), respectively. These parameters are generated randomly based on different distributions and cross correlations given in \cite[Tables 1-3]{Nokia}. The vector $\mathbf{a}(\phi,\theta)$ is the antenna array response vector, and it can considered as either uniform linear array (ULA) or uniform planar array (UPA). In order to enable beamforming in the elevation direction (3D beamforming), we use the uniform $U\times V$ planar array given by \cite{SS}
\begin{align}
\mathbf{a}(\phi,\theta)=\big[ 1, ..., e^{jkd_{\textrm{a}}(u\sin(\phi)\sin(\theta)+v\cos(\theta))}, ..., \nonumber \\ e^{jkd_{\textrm{a}}((U-1)\sin(\phi)\sin(\theta)+(V-1)\cos(\theta))} \big]^T
\end{align}
where $d_a$ is the distance between antenna elements, and $u\in\{1, ..., U\}$ and $v\in\{1, ..., V\}$ are the indices of antenna elements.

We consider two link states for each channel, (line of sight) LoS and (non-line of sight) NLoS, and use the following probability functions obtained based on the New York City measurements in \cite{RapLetter}
\begin{align}
p_{\text{LoS}}(d)&=\Big[\min\Big(\frac{d_\textrm{BP}}{d},1\Big).\Big(1-e^{-\frac{d}{\eta}}\Big)+e^{-\frac{d}{\eta}} \Big]^2\\
p_{\text{NLoS}}(d)&=1-p_{\text{LoS}}(d)
\end{align}
where $d$ is the 3D distance between UE and BS in meters, $d_{\textrm{BP}}$ is the breakpoint distance at which the LoS probability is not equal to 1 anymore, and $\eta$ is a decay parameter. The obtained values for these parameters are $d_{\textrm{BP}}=27$ m and $\eta=71$ m.

Moreover, we use the following omni-directional path loss model for LoS and NLoS links \cite{Nokia}
 % Sigma_SF = 4.9*randn Rappaport's paper Eq. (1) LoS
\begin{align}
PL[\textrm{dB}]=20\log_{10}\Big(\frac{4\pi d_0}{ \lambda}\Big) + 10n \log_{10}\Big(\frac{d}{d_0}\Big) + X_{\sigma_{\textrm{SF}}}
\end{align}
where $\lambda$ is the wavelength, $d_0$ is the reference distance, $n$ is the path loss exponent, and 
$X_{\sigma_{\textrm{SF}}}$ is the lognormal random variable with standard deviation $\sigma_{\textrm{SF}}$ (dB) which describes the shadow fading.
At 73 GHz, the path loss exponents and the shadowing factors are $n_{\textrm{LoS}}=2$, $n_{\textrm{NLoS}}=3.4$, $\sigma_{\textrm{SF, LoS}}=4.8$ dB, and $\sigma_{\textrm{SF, NLoS}}=7.9$ dB.
\subsection{System Model}
We consider a downlink scenario in a cellular mmWave MIMO network with $J$ BSs and $K$ UEs. $M_j$ and $N_k$ are the number of antennas at BS $j$ and UE $k$, respectively. Also, we assume $M_j \geq N_k$, which is a reasonable assumption. Let $\mathcal{J}=\{1, ..., J\}$ denotes the set of BSs and $\mathcal{K}=\{1, ..., K\}$ represents the set of UEs. Here, we consider TDD operation and assume that the channel state information (CSI) is available at both the transmitter and the receiver.
Each UE $k$ aims to receive $n_k$ data streams from its serving BS such that $1\leq n_k\leq N_k$, where the second inequality comes from the fact that the number of data streams for each UE cannot exceed the number of its  antennas. Thus, we can define the total number of downlink data streams sent by BS $j$ as
\begin{equation}
D_j=\sum_{k \in \mathcal{Q}_j(t)}n_k
\end{equation}
where $\mathcal{Q}_j(t)$ is called the \textit{Activation Set} and it represents the set of active UE in BS $j$ within time slot $t$, such that  $\mathcal{Q}_j(t) \subseteq \mathcal{K}$ and $|\mathcal{Q}_j(t)|=Q_j(t)\leq K$. Note that the total number of downlink data streams sent by each BS should be less than or equal to its number of antennas, i.e., $D_j \leq M_j$. For notational simplicity, we drop the time index $t$ in definition of $D_j$, and only keep the time index for $Q_j(t)$ due to its importance.

%%%%%%%%%%%%%%%%%%%%%%%%%%%%%%%%%%%%%%%%%%%%%%%
The $M_j\times 1$ transmitted signal from BS $j$ is given by
\begin{equation}\label{x_j}
\mathbf{x}_j = \mathbf{F}_j \mathbf{d}_j = \sum_{k\in \mathcal{Q}_j(t)}\mathbf{F}_{k,j}\mathbf{s}_k
\end{equation}
where $\mathbf{s}_k\in \mathbb{C}^{n_k}$ is the data stream vector for UE $k$ consists of mutually uncorrelated zero-mean symbols, with $\mathbb{E}\lbrack \mathbf{s}_k\mathbf{s}_k^*\rbrack = \mathbf{I}_{n_k}
$. The column vector $\mathbf{d}_j\in \mathbb{C}^{D_j}$ represents the vector of data symbols of BS $j$ which is the concatenation of the data stream vectors $\mathbf{s}_k,~k\in\mathcal{Q}_j(t)$, such that $\mathbb{E}\lbrack \mathbf{d}_j\mathbf{d}_j^*\rbrack = \mathbf{I}_{D_j}$.
$\mathbf{F}_{k,j}\in\mathbb{C}^{M_j\times n_k}$ is the linear precoder matrix that should be designed for each UE $k$ associated with BS $j$, and $\mathbf{F}_j\in\mathbb{C}^{M_j \times D_j}$ is the total linear precoder matrix of BS $j$ which is the concatenation of all $\mathbf{F}_{k,j}, k\in\mathcal{Q}_j(t)$.

The power constraint at BS $j$ can be described as 
\begin{equation}\label{power}
\mathbb{E}[\mathbf{x}_j^* \mathbf{x}_j]=\sum_{k\in \mathcal{Q}_j(t)}\textrm{Tr}(\mathbf{F}_{k,j}\mathbf{F}_{k,j}^*)\leq P_j 
\end{equation}
where $P_j$ is the transmit power of BS $j$.

Now, we can express the $N_k\times 1$ received signal at UE $k$ antennas as
\begin{equation}\label{y_k}
\mathbf{y}_k = \sum_{j\in \mathcal{J}}\mathbf{H}_{k,j}\mathbf{x}_j + \mathbf{z}_k
\end{equation}
and the final processed signal received by each UE is
\begin{equation}\label{y_tilde_k}
\tilde{\mathbf{y}}_k = \sum_{j\in \mathcal{J}}\mathbf{W}_k^* \mathbf{H}_{k,j}\mathbf{x}_j + \mathbf{W}_k^*\mathbf{z}_k
\end{equation}
where $\mathbf{W}_k\in\mathbb{C}^{N_k \times n_k}$ is the linear combiner matrix of UE $k$, $\mathbf{H}_{k,j}\in\mathbb{C}^{N_k\times M_j}$ represents the channel matrix between BS $j$ and UE $k$, and $\mathbf{z}_k\in\mathbb{C}^{N_k}$ is the white Gaussian noise vector at UE $k$, with $\mathbf{z}_k\sim\mathbb{CN}(\mathbf{0},N_0 \mathbf{I}_{N_k})$.
It is worth mentioning that in MIMO mmWave systems hybrid (analog and digital) beamforming should be implemented to reduces cost and power consumption of large antenna arrays \cite{SS}.
\begin{comment}
When UE $k$ is connected to BS $j$, the variance of desired received signal at user $k$ can be expressed as
\begin{align}\label{B}
\mathbf{G}_{k,j}=\mathbf{W}_{k}^*\mathbf{H}_{k,j}\mathbf{F}_{k,j} \mathbf{F}_{k,j}^* \mathbf{H}_{k,j}^*\mathbf{W}_{k}
\end{align}
Similarly, we can define the interference coming frm BS $i$ (while serving user $l$) to user $k$ as
\begin{align}\label{X}
\mathbf{X}_{l,i,k} &= \mathbf{W}_{k}^*\mathbf{H}_{k,i}\mathbf{F}_{l,i} \mathbf{F}_{l,i}^* \mathbf{H}_{k,i}^*\mathbf{W}_{k} 
\end{align}
In the next section, we will use this interference in formulating the network sum rate.
\end{comment}
\section{Time-Fractional User Association}
In the literature, when computing the instantaneous rate for a specific UE (connected to a BS), the interference coming from other UE-BS connections is assumed to be both independent of user association and present all the time (full interference). This assumption is not realistic and results in lower instantaneous user rates. In fact, network interference structure highly depends on user association and we need to consider this while computing the user rates. Moreover, user association depends on channel realizations which vary fast in mmWave frequencies. Thus, we cannot use the full interference structure in mmWave systems. 

In this section, we introduce a new user association model named \textit{Time-Fractional Association} (TFA). First, we need to introduce our definition of \textit{time slot} throughout this paper. Each time slot $t$ is a fraction of time which is considered to be comparable to channel coherence time such that the small-scale fading characteristics of the channel remains constant within it, and they only change from one time slot to another.
During time slot $t$ each UE is connected to one BS. Thus, the interference structure in each time slot depends on the user association on that specific slot. This interference structure is appropriate for mmWave channels where the channel variation can be very fast.
Moreover, we assume it is possible to split the data streams of each UE and transmit them in different time slots. 
Considering the above definitions, we study two association approaches in this paper: (i) instantaneous user association, which is performed within each time slot and results in unique association (each UE can be associated with only one BS during each time slot), and (ii) fractional (joint) user association, which is obtained by averaging over $T$ time slots. For each time slot, the mmWave channels are generated independently based on the channel model presented in Section II-A. We consider both approaches to evaluate the performance of our proposed TFA method and compare it with existing user association schemes.

We start by defining the \textit{Activation Matrix} as
\begin{equation}
\mathbf{B}\triangleq
\left[ \begin{array}{ccc}
\mathbold{\beta}(1)&\cdots &\mathbold{\beta}(T) 
\end{array}
\right]=
\left[ \begin{array}{ccc}
\beta_1(1) & \cdots & \beta_1(T) \\
\vdots & \ddots &\vdots \\
\beta_K(1) & \cdots & \beta_K(T) \\
\end{array} \right]
\end{equation}
where $\mathbold{\beta}(t)$ is called the \textit{Activation Vector} at time slot $t$, and each element of $\mathbf{B}$ is the index of BS to whom user $k$ is associated with during time slot $t$, i.e., $\beta_k(t)\in\mathcal{J}$ with $k\in\mathcal{K}$ and $t\in\mathcal{T}=\{1, ..., T\}$.  
Considering the above definition, the relationship between the activation set of BS $j$ and elements of activation matrix can be described as
\begin{equation}\label{Q_j}
\mathcal{Q}_j(t) = \{ k: \beta_k(t)=j\}.
\end{equation}
As stated earlier, we assume each UE can be associated with only one BS at any time slot $t$, i.e.,
\begin{equation}
\mathcal{Q}_j(t) \cap \mathcal{Q}_i(t) = \varnothing,~~j\neq i
\end{equation}
\begin{equation}\label{union}
\bigcup\limits_{j=1}^{J}\mathcal{Q}_j(t) = \mathcal{K}
\end{equation}
where (\ref{union}) follows from the fact that during each time slot, all UEs are served by networks' BSs.

The elements of activation matrix should satisfy the following conditions
\begin{align}
\sum_{j\in\mathcal{J}} 1_{\beta_k(t)}(j) &\leq 1, ~~\forall k\in \mathcal{K}\label{TFA_cons_1}\\
\sum_{k\in\mathcal{K}}1_{\beta_k(t)}(j). n_k &\leq D_j, ~~\forall j\in \mathcal{J}\label{TFA_cons_2}
\end{align}
where the indicator function is defined as 
\begin{align}
1_{\beta_k(t)}(j)= 
\begin{cases}
   1& ~~\textrm{if}~~ \beta_k(t)=j\\
   0& ~~\textrm{if}~~ \beta_k(t)\neq j
\end{cases}
\end{align}
The activation constraints in (\ref{TFA_cons_1}) reflect the fact that each UE cannot be associated with more than one BS in each time slot, and the resource allocation constraints in (\ref{TFA_cons_2}) denote that sum of data streams of UEs served by each BS cannot exceed the total number of available data streams on that BS. Note that $1_{\beta_k(t)}(j)$ is equal to one only if $\beta_k(t)=j$ or equivalently, $k\in\mathcal{Q}_j(t)$. Thus, the summation in (\ref{TFA_cons_2}) is actually over the set of active users in BS $j$.

Now, we define the \textit{Association Matrix} $\mathbf{A}$ as follows
\begin{equation}
\mathbf{A}\triangleq
\left[ \begin{array}{ccc}
\alpha_{1,1} & \cdots & \alpha_{1,J} \\
\vdots & \ddots &\vdots \\
\alpha_{K,1} & \cdots & \alpha_{K,J} \\
\end{array} \right]
\end{equation}
where $\alpha_{k,j}\in [0,1]$ is the association coefficient (fraction), and it represents the average connectivity of UE $k$ to BS $j$. If $\alpha_{k,j}=0$, we say UE $k$ is not associated with BS $j$. 
In this model, association coefficients are considered as a fraction of time. More specifically, we assume that $\alpha_{k,j}$ is the average fraction of time during which UE $k$ is connected to BS $j$. 

The relationship between the association coefficients and the elements of activation matrix is given by
\begin{equation}\label{alpha_beta}
\alpha_{k,j} = \lim_{T\rightarrow \infty} \frac{1}{T}\sum_{t=1}^T 1_{\beta_k(t)}(j)
\end{equation}
According to (\ref{alpha_beta}), given the activation matrix $\mathbf{B}$, one can easily obtain the association matrix $\mathbf{A}$. 

\section{User Association Optimization Problem}
In this section, we evaluate the instantaneous and average per-user throughputs by processing the received signal at each user. Then, we formulate an optimization problem and use a heuristic search method to find the optimal user association.
\subsection{Formulation of instantaneous user rate}
Considering (\ref{x_j}) and (\ref{y_tilde_k}), the signal received by UE $k$ at slot $t$ can be decomposed as
\begin{align}\label{y_tilde_k_t}
\tilde{\mathbf{y}}_k (t)&= \sum_{j\in \mathcal{J}}\mathbf{W}_k^* \mathbf{H}_{k,j} \mathbf{x}_j + \mathbf{W}_k^* \mathbf{z}_k\nonumber \\
&=  \underbrace{\mathbf{W}_k^*\mathbf{H}_{k,j}\mathbf{F}_{k,j} \mathbf{s}_k}_\text{Desired signal} + \underbrace{\mathbf{W}_k^*\mathbf{H}_{k,j}\sum_{\substack{l\in \mathcal{Q}_j(t) \\ l\neq k}}\mathbf{F}_{l,j} \mathbf{s}_l}_\text{Intra-cell interference} \nonumber\\
&+ \underbrace{\mathbf{W}_k^*\sum_{\substack{i\in \mathcal{J} \\ i\neq j}}\sum_{\substack{l\in \mathcal{Q}_i(t)}} \mathbf{H}_{k,i} \mathbf{F}_{l,i} \mathbf{s}_l}_\text{Inter-cell interference} + \underbrace{\mathbf{W}_k^*\mathbf{z}_k}_\text{Noise}
\end{align}
where the first term is the received signal from desired BS ($j$), the second term represents the interference coming from the same BS ($j$) by signals intended for its other active UEs, the third term is the interference coming from other BSs ($i\neq j$) by signals sent to their active UEs, and the last term is the received noise at UE $k$.
The activation sets $\mathcal{Q}_j(t)$ and $\mathcal{Q}_i(t)$ appeared in the interference terms indicate that the interference highly depends on the user association, which highlights the novelty of this work.
Again, we note that all vectors and matrices in (\ref{y_tilde_k_t}) are time-dependent, and the time index $t$ is dropped for the sake of notational simplicity. 

When UE $k$ is connected to BS $j$ in time slot $t$, its instantaneous rate can be obtained as \cite{Telatar}
\begin{equation}\label{R_kj}
R_{k,j}(t) = \log_2\Big |\mathbf{I}_{N_k} + (\mathbf{Y}_{k,j}(t))^{-1}\mathbf{W}_{k}^*\mathbf{H}_{k,j}\mathbf{F}_{k,j} \mathbf{F}_{k,j}^* \mathbf{H}_{k,j}^*\mathbf{W}_{k}\Big |
\end{equation}
\begin{align}
\mathbf{Y}_{k,j}&(t)= \mathbf{W}_{k}^*\mathbf{H}_{k,j}\Big( \sum_{\substack{l\in \mathcal{Q}_{j}(t) \\ l\neq k}} \mathbf{F}_{l,j} \mathbf{F}_{l,j}^* \Big ) \mathbf{H}_{k,j}^*\mathbf{W}_{k}  \nonumber \\
&+ \mathbf{W}_{k}^* \Big( \sum_{\substack{i\in \mathcal{J} \\ i\neq j}} \sum_{\substack{l\in \mathcal{Q}_i(t)}} \mathbf{H}_{k,i}\mathbf{F}_{l,i} \mathbf{F}_{l,i}^* \mathbf{H}_{k,i}^* \Big ) \mathbf{W}_{k} + N_0 \mathbf{W}_k^*\mathbf{W}_k
\end{align}
\begin{comment}
Considering (\ref{B}) and (\ref{X}), we can rewrite the instantaneous rate in compact form as
\begin{equation}\label{inst_rate_t}
R_{k,j}(t) = \log_2\det (\mathbf{I}_{N_k} + (\mathbf{Y}_{k,j}(t))^{-1}\mathbf{G}_{k,j})
\end{equation}
\begin{align}
\mathbf{Y}_{k,j}(t) &= \sum_{\substack{l\in \mathcal{Q}_{j}(t) \\ l\neq k}} \mathbf{X}_{l,j,k} + \sum_{\substack{i\in \mathcal{J} \\ i\neq j}} \sum_{\substack{l\in \mathcal{Q}_i(t)}} \mathbf{X}_{l,i,k} + N_0 \mathbf{W}_k^*\mathbf{W}_k
\end{align}
\end{comment}
The instantaneous rate given in (\ref{R_kj}) is a function of activation sets $\mathcal{Q}_j(t)$. Thus, the instantaneous per-user throughput at time slot $t$ can be expressed as 
\begin{align}\label{r_k_t}
r_{k}(t)=\sum_{j\in\mathcal{J}} 1_{\mathbold{\beta}_k(t)}(j)\times R_{k,j}(t)
\end{align}
and the average per-user throughput is given by
\begin{align}
r_{k}=\lim_{T\rightarrow \infty}\frac{1}{T}\sum_{t=1}^{T} r_k(t)
\end{align}
\subsection{Optimization Problem}
As stated before, the channel variation can be very fast in mmWave frequencies and the small-scale characteristics of the channel could change a lot even during two consecutive time slots. Thus, we need to perform the user association in each time slot.
Defining the instantaneous user throughput vector $\mathbf{r}(t)\triangleq (r_1(t), ..., r_K(t))$, we wish to find the optimal activation vector $\mathbold{\beta}(t)$ which maximizes an overall network utility function. This utility function should be concave and monotonically increasing. 
In this paper, we consider the well-known and widely used sum-rate utility function defined by 
%Thus, we consider the well-known and widely used family of utility functions defined by
\begin{align}\label{u(r(t))}
U(\mathbf{r}(t))&=\sum_{k\in\mathcal{K}} r_k(t)=\sum_{k\in\mathcal{K}}\sum_{j\in\mathcal{J}} 1_{\mathbold{\beta}_k(t)}(j)\times R_{k,j}(t)
\end{align}
Thus, the optimization problem for each time slot $t$ can be written as
\begin{subequations}\label{opt_prob2}
\begin{align}
\maxi_{\beta_k(t)\in \mathcal{J}}~~&U(\mathbf{r}(t))\\
\mathrm{subject~to}~~&\sum_{j\in\mathcal{J}} 1_{\mathbold{\beta}_k(t)}(j) \leq 1, ~~\forall k\in \mathcal{K}\\
&\sum_{k\in\mathcal{K}}1_{\mathbold{\beta}_k(t)}(j). n_k \leq D_j, ~~\forall j\in \mathcal{J}
\end{align}
\end{subequations}
%%%%%%%%%%%%%%%%%%%%%%%%%%%%%%%%%%%%%%%%%
This is an optimization problem with the integer variables $\beta_k(t)\in\mathcal{J}$ for $k\in\mathcal{K}, t\in\mathcal{T}$. Here, we use equal power allocation to split the BS power among its active users. Thus, the power constraint in (\ref{power}) is no longer applicable and can be ignored.

In this paper, we use singular value decomposition (SVD) to obtain the precoder and combiner matrices (SVD beamforming). To this end, we first need to decompose the channel matrix $\mathbf{H}\in\mathbb{C}^{N_k\times M_j}$ as
\begin{align}
\mathbf{H} &= \mathbf{\Phi}\mathbf{\Sigma}\mathbf{\Gamma}^*
\end{align}
where $\mathbf{\Phi}\in\mathbb{C}^{N_k\times \textrm{rank}(\mathbf{H})}$ is the unitary matrix of left singular vectors, $\mathbf{\Sigma}\in\mathbb{C}^{\textrm{rank}(\mathbf{H})\times \textrm{rank}(\mathbf{H})}$ is the diagonal matrix of singular values (in decreasing order), and $\mathbf{\Gamma}\in\mathbb{C}^{M_j\times \textrm{rank}(\mathbf{H})}$ is the unitary matrix of right singular vectors. Then, we partition the channel matrix as 
\begin{align}\label{Ch_partitioning}
\mathbf{H} &= 
\left[
\begin{matrix}
\mathbf{\Phi}_1 & \mathbf{\Phi}_2
\end{matrix}
\right]
\left[
\begin{matrix}
\mathbf{\Sigma}_1 &  \mathbf{0}\\
\boldsymbol{0} &  \mathbf{\Sigma}_2
\end{matrix}
\right]
\left[
\begin{matrix}
\mathbf{\Gamma}_1^* \\ \mathbf{\Gamma}_2^*
\end{matrix}
\right]\nonumber \\
&= \mathbf{\Phi}_1\mathbf{\Sigma}_1\mathbf{\Gamma}_1^*+ \mathbf{\Phi}_2\mathbf{\Sigma}_2\mathbf{\Gamma}_2^*
\end{align}
where  $\mathbf{\Phi}_1\in\mathbb{C}^{N_k \times n_k}$, $\mathbf{\Sigma}_1\in\mathbb{C}^{n_k \times n_k}$, $\mathbf{\Gamma}_1\in\mathbb{C}^{M_j \times n_k}$, and $n_k$ is the number of data streams intended for user $k$.
The above partitioning is done to extract the precoder and combiner of appropriate sizes. More specifically, each precoder $\mathbf{F}_{k,j}$ and combiner $\mathbf{W}_k$ need to be of size $M_j\times n_k$ and $N_k\times n_k$, respectively. 
Then, the SVD precoder and combiner can be obtained as %\cite{SS}
\begin{align}
\mathbf{F}&=\mathbf{\Gamma}_1\\
\mathbf{W}&=\mathbf{\Phi}_1
\end{align}

\begin{figure}
\centering
\hspace*{-1.7em}
\includegraphics[scale=.25]{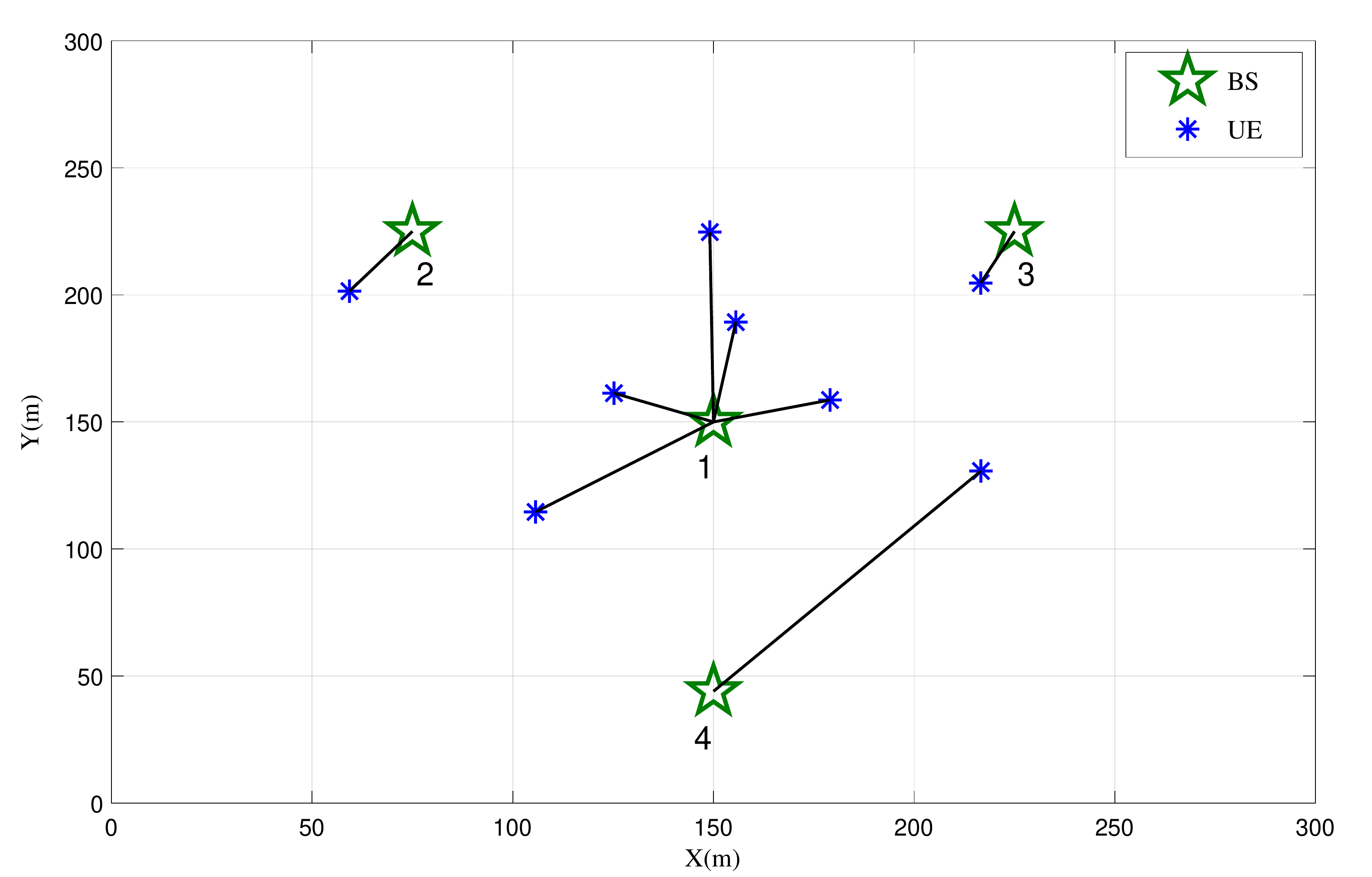}
   \vspace*{-1.1em}
\caption{Max-SINR user association with full interference}
\label{max-SINR}
\end{figure}

The optimization problem in (\ref{opt_prob2}) is a mixed integer nonlinear programming (MINLP), which is known to be NP-hard due to its nonlinear structure and presence of integer variables. The nonlinearity comes from the indicator function appeared in (\ref{r_k_t}) and constraints (\ref{opt_prob2}b-c).
These problems are typically difficult to solve due to their combinatorial structure and potential presence of multiple local minima in the search space. 

Genetic algorithms (GA) are considered as powerful and effective tools for solving combinatorial optimization problems.
GA is a method based on natural selection which simulates biological evolution. The algorithm iteratively generates and modifies a population of individual solutions. 
After successive generations, the population eventually evolves toward an optimal solution \cite{GA}.
In the next section, we use the GA solver provided in Global Optimization Toolbox of MATLAB to solve the optimization problem in (\ref{opt_prob2}). 
\section{Numerical Results}
In this section, we investigate the performance of the proposed user association scheme in a simple mmWave MIMO network operating at 73 GHz with $J=4$ BSs and $K=8$ UEs. The mmWave links are generated as described in Section II-A, and each link is composed of 5 clusters with 10 rays per cluster. Each BS is equipped with a $8\times 8$ UPA of antennas, and each UE is equipped with a $2\times 2$ UPA of antennas. The Noise power spectral density is $-174$ dBm/Hz, and all BSs transmit at the same power level $P_j$.
Moreover, we assume that the network nodes are deployed in a region of size $300~\textrm{m} \times 300~\textrm{m}$. The BSs are placed at specific locations and the UEs are distributed randomly within the given area, as shown in Fig. \ref{max-SINR}. There are $n_k=2$ data streams for each UE and the total number of data streams sent by each BS is $D_j=4$. Thus, the maximum number of allowed active users at each BS is $Q_j(t)=2$, and a BS is considered to be overloaded (congested) if more than 2 UEs are associated with it.
\begin{figure}
\centering
\hspace*{-1.7em}
\includegraphics[scale=.25]{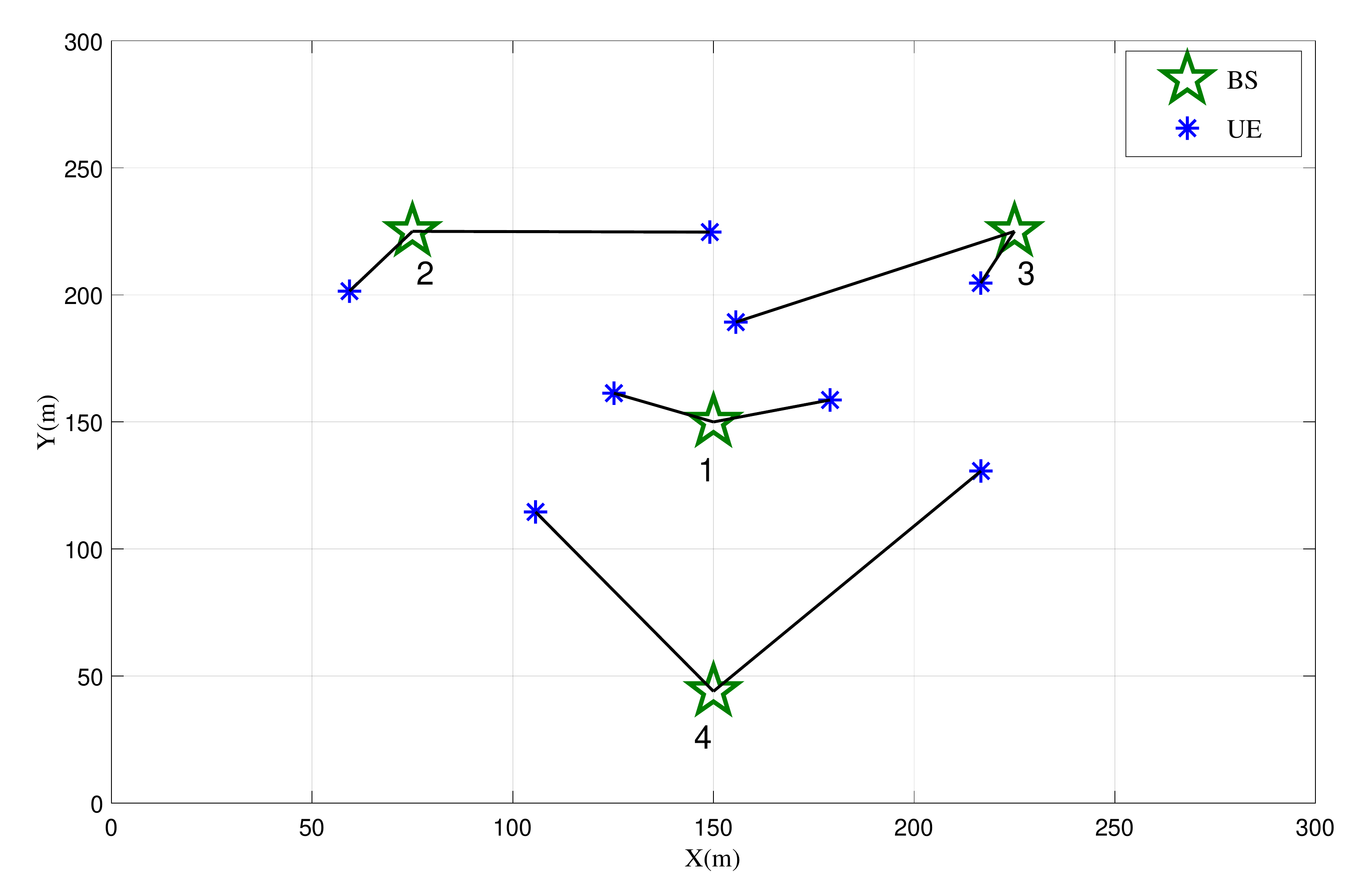}
   \vspace*{-1.1em}
\caption{Proposed load balancing TFA scheme}
\label{TFA_fig}
\end{figure}

First, we compare the TFA model with the conventional max-SINR scheme. 
Fig. \ref{max-SINR} shows the result of max-SINR association with full interference, where the BS at the center of the network (BS 1) is overloaded by 3 extra UEs. 
Load balancing user association using the proposed TFA scheme is shown in Fig. \ref{TFA_fig}. It can be seen from the figure that the proposed method perfectly balances the BSs' load by pushing the overloading UEs from the congested BS to other BSs. 

\begin{figure*}
   \centering
\hspace*{-3em}
   \includegraphics[scale=.35]{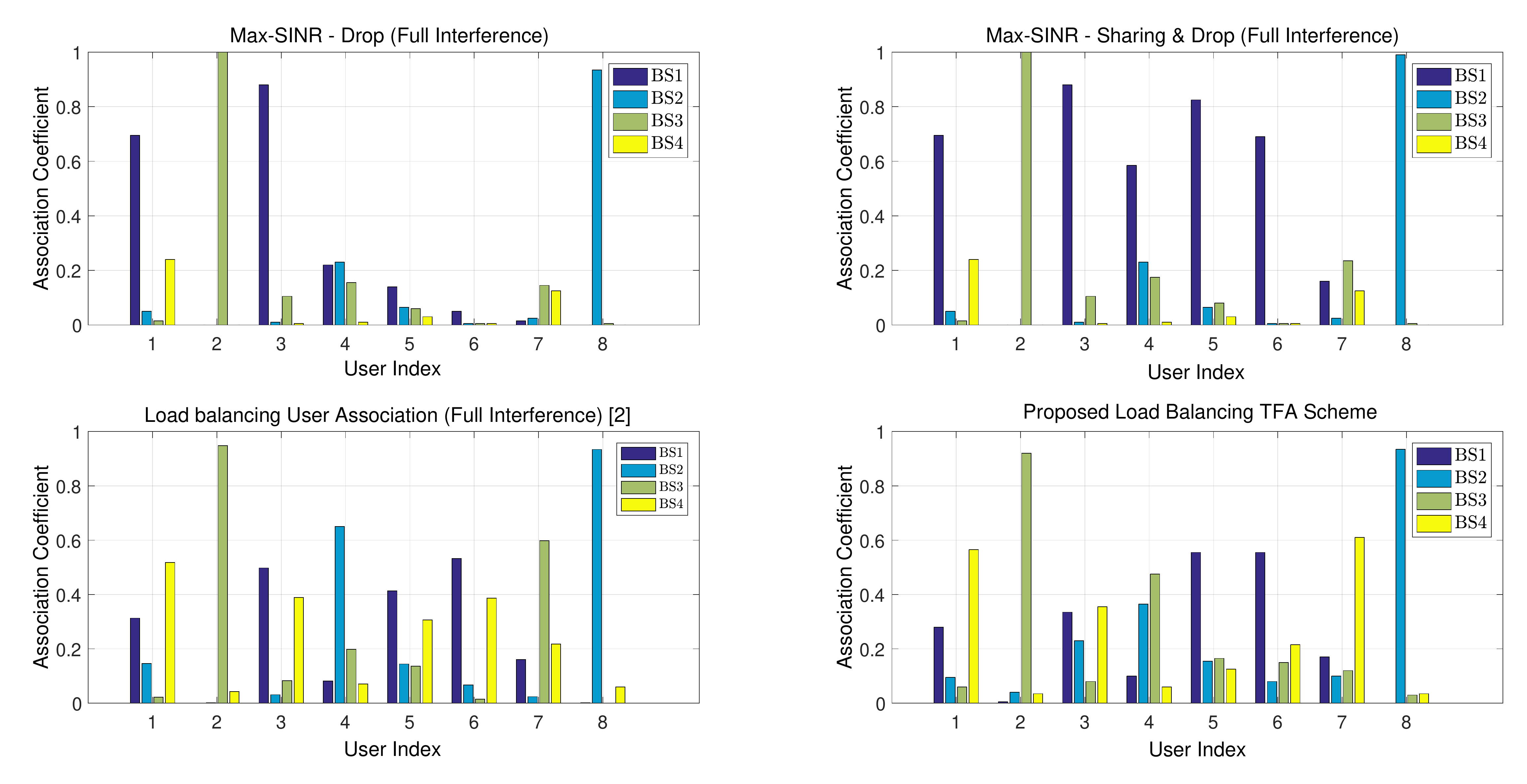}
   \vspace*{-1.5em}
   \caption{Comparison of association coefficients for different user association schemes}
   \label{Ass_Coeff}
\end{figure*}

Next, we compare the performance of the TFA method with three other user association schemes: (i) Max-SINR - Drop, (ii) Max-SINR - Sharing \& Drop, and (iii) Load Balancing User Association in \cite{Caire}. In the first method, those UEs who overloaded the congested BS are dropped, and in the second one, the data streams of the congested BS are shared among maximum number of UEs that it can serve. For instance, in our scenario depicted in Fig. \ref{max-SINR}, the available 4 data streams of BS 1 are shared among the first 4 UEs (which receive the highest SINR from BS 1) and the 5th UE is dropped. For the last scheme, we perform the load balancing user association based on the approach presented in \cite{Caire}. 
Fig. \ref{Ass_Coeff} compares the association coefficients (averaged over 1000 time slots) for above schemes. It is clear from the figure that both load balancing user association schemes successfully balance the BSs' loads.

\begin{figure}
\centering
\hspace*{-1.3em}
\includegraphics[scale=.2835]{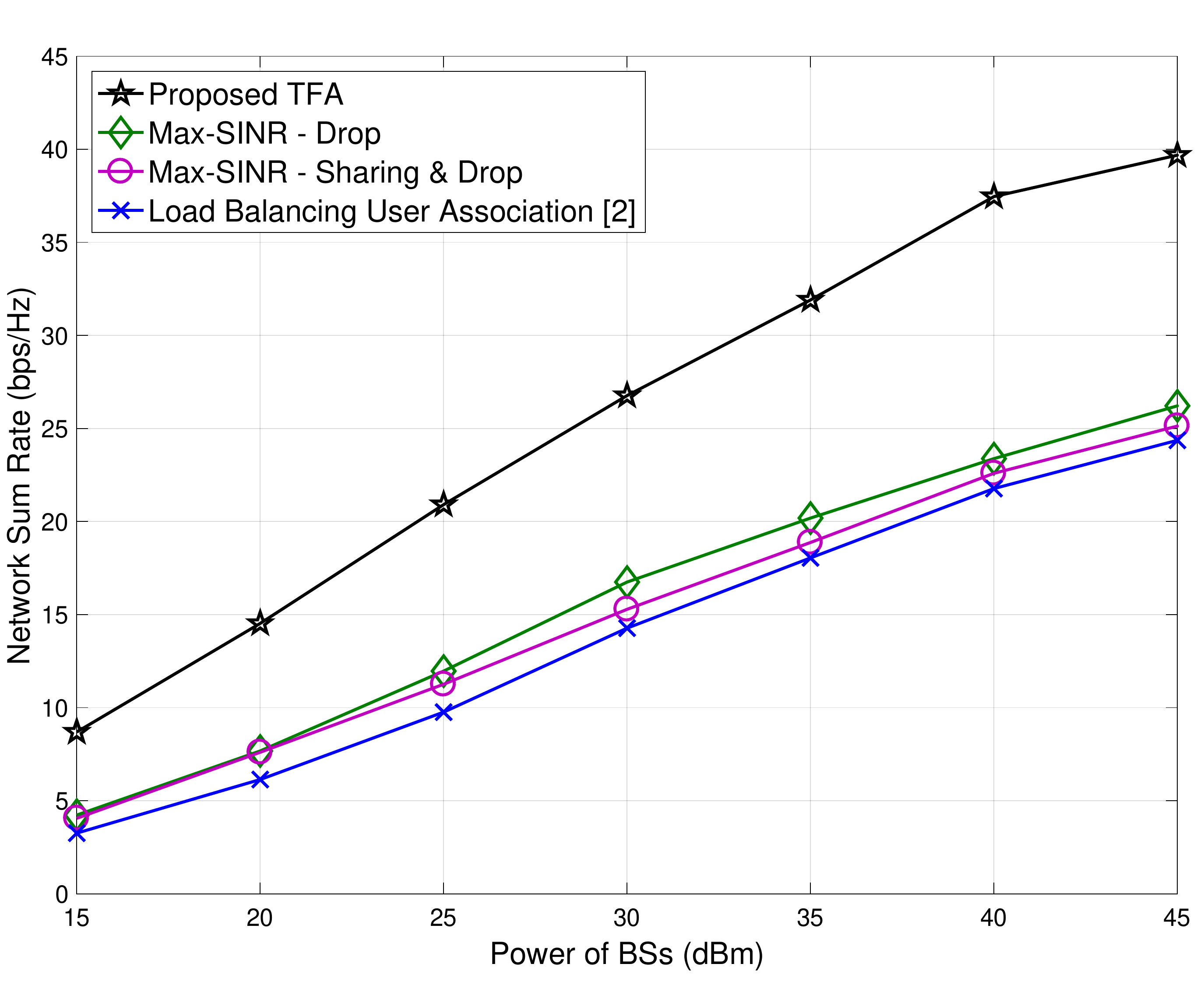} %348
   \vspace*{-1.2em}
\caption{Network sum rate for user association schemes in a network with $4$ BSs, $8$ UEs, $8\times8$ UPA at each BS, and $2\times2$ UPA at each UE. BSs' locations are as in shown Fig. 1, and UEs are distributed randomly in an area of $300\times300$ $\textrm{m}^2$. The BSs transmit at the same power level, and the noise power spectral density is $-174$ dBm/Hz.}
\label{SumRate}
\end{figure}

Finally, we examine the performance of the TFA scheme in terms of network sum rate. 
Fig. \ref{SumRate} depicts the network sum rate (given in (\ref{u(r(t))})) versus the BSs' transmit power for different association schemes. 
Note that all other three association schemes assume a full interference structure. It can be inferred from the figure that network interference highly depends on user association, since our TFA method outperforms the other user association schemes which all ignore the effect of user association on the network interference. Also, we can see that the load balancing scheme presented in \cite{Caire} slightly underperforms the max-SINR schemes.
This result is expected since, contrary to max-SINR schemes, the load balancing approach takes into account the BS loads.

\section{Conclusion}
In this paper we investigated the problem of optimal user association in a mmWave MIMO network. 
We first introduced the activation matrix and showed that the user instantaneous rate is a function of the elements of this matrix. 
Then, we formulated a new association model, called TFA, in which network interference depends on user association. The performance of the proposed TFA scheme is investigated by considering three other association schemes: (i) max-SINR with user drop, (ii) max-SINR with resource sharing and user drop, and (iii) load balancing user association proposed in [2]. Simulation results confirmed the fact that the network interference structure highly depends on user association and showed that the proposed scheme outperforms all other three methods, which ignore the effect of interference on user association.
%***********************************************************************

\vspace*{-10em}

\end{document}